\documentclass[final,3p,times,twocolumn]{elsarticle}

\usepackage{graphicx}
\usepackage{amssymb}
\usepackage{amsmath}
\usepackage{xcolor}
\usepackage{lineno}

\journal{Nuclear Instruments and Methods, Section A}

\begin{document}

\begin{frontmatter}


\title{Surface mechanisms governing long-term stability \\of GEM detectors in CO$_2$-based gaseous mixtures}



\author[IFUSP]{Tiago F. Silva\fnref{myfootnote}}
\fntext[myfootnote]{Corresponding author. e-mail: tfsilva@if.usp.br}
\author[IFUSP]{Thiago B. Saramela}
\author[IFUSP]{Willian W.R.A. da Silva}
\author[UFRJ]{Camilla de S. Codeço}
\author[IQUFRGS]{\\Maria do C. M. Alves}
\author[IFUFRGS]{Jonder Morais}
\author[IPEN]{Niklaus U. Wetter}
\author[IPEN]{Anderson Z. de Freitas}

\address[IFUSP]{Instituto de Física da Universidade de S\~ao Paulo, Rua do Matão, 1371, 05508-090 S\~ao Paulo, Brazil.}
\address[UFRJ]{
Instituto de Física da Universidade Federal do Rio de Janeiro, Av. Athos da Silveira Ramos, 149, 21941-909, Rio de Janeiro, Brazil.}
\address[IQUFRGS]{Instituto de Química da Universidade Federal do Rio Grande do Sul,  Av. Bento Gonçalves, 9500, 91501-970, Rio Grande do Sul, Brazil.}
\address[IFUFRGS]{Instituto de Física da Universidade Federal do Rio Grande do Sul, Av. Bento Gonçalves, 9500, 91501-970, Rio Grande do Sul, Brasil}
\address[IPEN]{Instituto de Pesquisas Energéticas e Nucleares, Av. Prof. Lineu Prestes 2242, 05508-000 São Paulo, Brazil}

\begin{abstract}
Understanding the chemical stability of Gas Electron Multipliers (GEMs) operated in CO$_2$-based mixtures is essential for improving detector longevity and reliability. In this work, we investigate the interaction between CO$_2$ molecules and the copper electrodes of GEM foils through near-ambient pressure X-ray photoelectron spectroscopy (NAP-XPS) and complementary Raman mapping. The measurements reveal that CO$_2$ exposure promotes a mild reduction of CuO to Cu$_2$O on untreated surfaces, while sputter-cleaned foils remain metallic and chemically stable. Raman spectroscopy confirms the predominance of Cu$_2$O with spatially heterogeneous contributions from CuO at the micrometer scale, providing structural support for the oxidation-state evolution inferred from XPS. Carbon 1s spectra identify carbonyl (C=O), C-O, carbonate, and hydroxyl species, indicating that oxidized copper sites mediate surface reactions and the formation of oxygenated films. A spectral feature consistent with ionized gas phase CO$_2$ species is observed in the O 1s region, suggesting that a fraction of the gas phase may become ionized in the near-surface region during acquisition. This is relevant for GEM detectors, where CO$_2^{+}$ and other ionized species generated in the avalanche can interact with the copper electrodes. These findings indicate that CO$_2$ acts not only as a quencher but also as a weakly reactive component capable of establishing self-limiting redox equilibria that favor the formation of thin, inorganic oxygenated layers. Such layers are expected to be significantly less prone to charge accumulation than the polymeric or carbonaceous deposits typically formed in hydrocarbon-based mixtures. The results provide experimental insight into the mechanisms underlying GEM stability and contribute to a deeper understanding of aging phenomena in GEM-based systems. 
\end{abstract}


\end{frontmatter}


\section{Introduction}
\label{intro}

    Gaseous detectors remain essential instruments for charged-particle tracking and radiation imaging in nuclear and high-energy physics. Their operation relies on electron multiplication processes in gas media, whose stability and reproducibility are key to long-term performance. However, sustained irradiation in the presence of quenching gases, organic impurities, and polymeric components often leads to progressive changes in gain, electronic noise, impacting operational stability and worsening energy resolution, a phenomenon generally referred to as aging. Since the early systematic studies by Sauli and Va’vra~\cite{sauli_gem:_1997, vavra_physics_2003, sauli_fundamental_2003} aging has been associated with the formation of thin insulating or semiconducting films on electrode surfaces, resulting from polymerization, oxidation, or chemical redeposition processes initiated during gas avalanches.
    
    In the context of Micro-Pattern Gaseous Detectors (MPGD), the Gas Electron Multiplier (GEM) introduced by Sauli in 1997~\cite{sauli_gem:_1997} represented a technological leap in robustness and radiation hardness. The confinement of avalanche processes within microscopic holes and the use of inert gas mixtures substantially mitigated the classical aging mechanisms known from wire chambers. Nevertheless, long-term studies have shown that GEMs are not entirely immune to degradation. Accelerated irradiation campaigns have revealed that specific conditions (such as gas impurities, surface oxides, or the presence of silicon-based contaminants) can promote the gradual formation of insulating films or chemical modifications at the copper–polyimide interface~\cite{niebuhr_aging_2006, fallavollita_advanced_2020, poli_lener_irradiation_2024}. These changes may alter the local field configuration, leading to charge-up phenomena and occasional discharges of the Malter type.
    
    Recent investigations have deepened the microscopic understanding of these processes. Our results~\cite{saramela_evidence_2024} provided direct spectroscopic evidence for the redeposition of polyimide fragments within the GEM holes and the migration of chromium from the inner metallic layers of the foil. These findings, supported by Time-Of-Flight Secondary Ion Mass Spectroscopy (ToF-SIMS) and high-resolution electron microscopy, confirmed that aging is not purely an electrical or mechanical effect, but also a chemical transformation of the electrode surface driven by the plasma environment. In parallel, other studies have shown that the chemistry of the gas environment plays a decisive role: reactive species derived from CF$_4$ or trace hydrocarbons tend to produce aggressive fluorinated or carbonaceous films, whereas the use of CO$_2$ as a quencher results in comparatively mild and reversible surface modifications~\cite{corbetta_triple-gem_2020, procureur_why_2020}. Tests with gas recirculation have further demonstrated that even trace levels of H$_2$O, O$_2$, or Si-containing molecules can drastically influence detector longevity, linking the problem of aging to the broader question of gas purity and material compatibility.
    
    Understanding how these factors converge to produce insulating layers on the GEM electrodes remains a critical challenge. The composition, morphology, and chemical state of these layers determine whether the surface behaves as a benign oxide, a resistive passivation, or an electrically active deposit capable of inducing discharges. The present work addresses this question by using Near-Ambient Pressure X-ray Photoelectron Spectroscopy (NAP-XPS) surface analysis to provide experimental evidence of how surface conditions control the formation of insulating films on copper electrodes. To complement the surface-sensitive chemical information obtained by NAP-XPS, Raman spectroscopy was employed to probe the structural distribution of copper oxide phases at the micrometer scale. Furthermore, it offers insights into why CO$_2$-based mixtures, despite supporting carbonate formation, exhibit markedly less aggressive aging behavior compared to hydrocarbon-quenched systems.

    Despite the widespread use of CO$_2$-based mixtures in GEM detectors, direct spectroscopic evidence of the initial surface chemistry between CO$_2$ molecules and the copper electrodes remains scarce. Understanding these early surface interactions is essential for clarifying why CO$_2$ mixtures exhibit significantly milder aging behavior compared with hydrocarbon-based gases. The present work addresses this gap by combining in-situ X-ray Photoelectron Spectroscopy and Raman spectroscopy to investigate the chemical processes occurring at the Cu–CO$_2$ interface.

\section{Methods}

    \subsection{Sample preparation and the NAP-XPS measurements}
    \label{sec:sample_prep}
    
\begin{figure*}[htb!]
            \centering
            \includegraphics[width=0.8\linewidth]{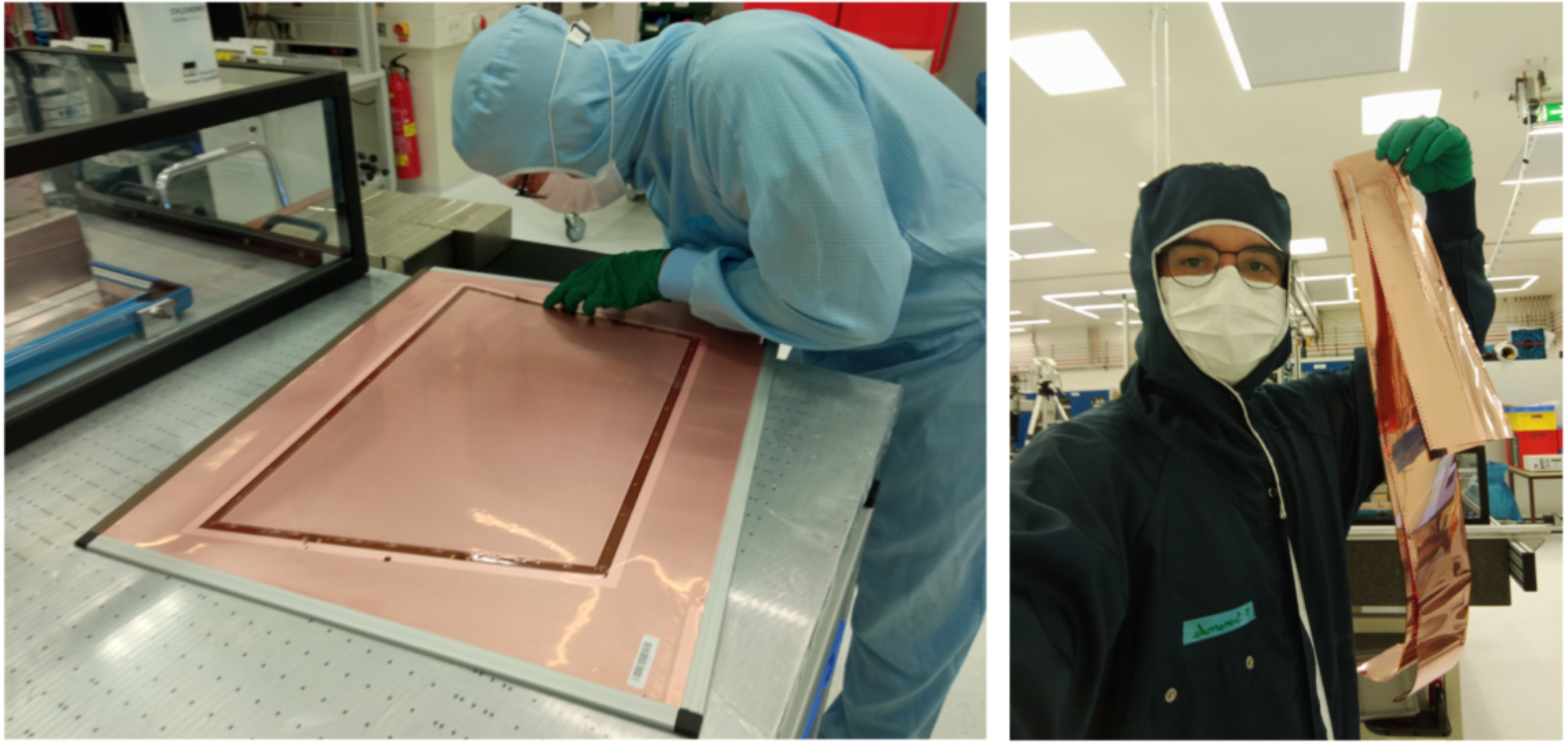}
            \caption{Sample preparation from the leftovers of ALICE-TPC spare parts construction.}
            \label{fig:samples}
        \end{figure*}

        The GEM samples analyzed in this work were produced at CERN and obtained from leftovers during the construction of spare parts for the Time Projection Chamber (TPC) of the ALICE experiment at CERN, with assembly activities carried out at GSI center in Darmstadt (see Fig. \ref{fig:samples}). These foils were stored under the same environmental conditions as those installed in the detector and had never been used in operation. Small pieces of approximately $1 \times 1~\text{cm}^2$ were manually cut from the unused foils using stainless steel scissors, ensuring minimal mechanical deformation.
        
        Two types of samples were investigated: as-received and cleaned. The as-received samples were analyzed without any pre-treatment to preserve the surface state representative of the original material. The cleaned sample was pre-treated to remove the surface contaminants and the native oxide layer while maintaining the integrity of the copper clad.
        
        NAP-XPS measurements and sample preparation were carried out in a UHV (ultra high vacuum) system of the Brazilian Center for Physical Research (CBPF). The UHV system is composed of an analysis chamber equipped with a NAP-XPS coupled to the preparation chamber, both with a base pressure of 2x$10^{-9}$ mbar. The cleaned sample was prepared by one sputtering cycle of a $Ar^+$ ion gun with 1 keV of energy for 40 min at 5 × $10^{-6}$ mbar of Ar pressure in the preparation chamber. The sample was then moved internally to the UHV system to the NAP-XPS chamber with no exposure to air.               

        The analysis chamber features a Phoibos spectrometer (SPECS) equipped with a monochromatic Al-$K{\alpha}$ X-ray source (photon energy of 1486.6~eV). Under these excitation conditions, the XPS technique exhibits an information depth typically in the range of 5–10~nm, depending on the material and emission angle~\cite{racz_evaluation_2025}. For these measurements, the samples were mounted on conductive sample holders using carbon adhesive tape to ensure good electrical grounding. The analysis chamber also has a leak valve that allows $CO_2$ to be inserted during XPS measurements. During in situ experiments at room temperature, the pressure of $CO_2$ varies from $10^{-6}$ mbar to 1 mbar. 

    \subsection{NAP-XPS Data analysis}

        The NAP-XPS data were processed using the CasaXPS software package~\cite{FAIRLEY2021100112}. 
        A Shirley-type nonlinear background~\cite{Shirley1972} was subtracted before curve fitting. All peaks were fitted using mixed Gaussian–Lorentzian (GL(30)) line shapes with fixed full width at half maximum (FWHM) across similar chemical states. 
        
        Binding energies were referenced to the main C–C/C–H component at 284.8~eV to correct for any potential surface charging effects. The C~1s, and Cu~2p regions were fitted according to literature assignments~\cite{favaro_subsurface_2017, Moulder1992}, with specific components representing C–C/C–H, C–O, C=O (carbonyl), and O–C=O (carbonate) in the C~1s region; and Cu$^0$, Cu$^+$, and Cu$^{2+}$ oxidation states in the Cu~2p region, including the Cu$^{2+}$ shake-up satellites. 

        Before the introduction of CO$_2$ into the chamber, the spectral components at approximately 287.9 eV and 288.4 eV should not be attributed to chemisorbed or physisorbed CO$_2$, since these species cannot exist under UHV conditions. Instead, they correspond to pre-existing oxygenated functionalities, such as C–O and C=O groups, originating from residual surface oxidation or adventitious carbon. Upon CO$_2$ exposure, these same binding energy ranges become representative of molecular adsorption, evolving respectively toward bridge-bonded (chemisorbed) and linear (physisorbed) CO$_2$ configurations.
        
        Quantitative elemental compositions were obtained using the relative sensitivity factors provided in the CasaXPS library. All data sets were treated using identical fitting parameters and constraints to ensure internal consistency. The resulting binding energies and relative intensities were then used to identify the chemical states of copper, oxygen, and carbon species present on the GEM surfaces under the various experimental conditions.

    \subsection{Raman Spectroscopy}

        Raman spectroscopy measurements were performed to identify and spatially resolve copper oxide phases present on the GEM foil surface. The analyzes were carried out using a LabRam HR Evolution spectrometer (HORIBA) equipped with a 785 nm excitation laser operated at a power of 24 mW. Under these excitation conditions, the effective information depth in metals is typically in the range of 10–20 nm, being primarily limited by the optical skin depth of the material. The optical configuration included a 50× long-working-distance objective (LWDIR), a 500 $\mu$ pinhole, and a 600 lines/mm diffraction grating. Spectra were acquired with an integration time of 5 s and 2 accumulations per measurement point. To evaluate the spatial distribution of oxide phases, a 30 × 30 point mapping was performed over selected surface regions.

        The Raman bands were analyzed to identify copper oxide phases based on characteristic vibrational modes. A band centered at approximately 150 cm$^{-1}$ was assigned to Cu$_2$O, while a band near 300 cm$^{-1}$ was attributed to CuO~\cite{zoolfakar_nanostructured_2014}. 

    \subsection{Expected behavior of CO$_2$ on copper surfaces}

        The interaction between CO$_2$ and copper surfaces is a multifaceted phenomenon governed by the oxidation state of copper, the presence of electric fields or excess charge, and the availability of surface and subsurface oxygen. Early NAP-XPS studies revealed that under pure CO$_2$ exposure, metallic Cu surfaces exhibit primarily weak physisorption, characterized by minor contributions from C–O(H) and carbonate-like species in the C 1s region~\cite{regoutz_influence_2018} and $CO_2^{\delta-}$ \cite{salmeron_2016, salmeron_2008}. The introduction of trace O$_2$ or humidity, however, leads to partial oxidation of the surface to Cu$_2$O, significantly enhancing the formation of oxygenated species such as monodentate and bidentate carbonates and bicarbonates \cite{salmeron_2008}. These findings indicate that the surface oxidation state critically modulates CO$_2$ activation and binding strength.
        
        Complementary thermogravimetric and XPS analyzes by Isahak \textit{et al.}~\cite{isahak_adsorptiondesorption_2013} confirmed that CO$_2$ adsorption is substantially stronger on CuO than on Cu$_2$O. The chemisorption of CO$_2$ on CuO yields copper carbonate species (CuCO$_3$) with decomposition temperatures near 583 K, while Cu$_2$O exhibits only weak physisorption~\cite{isahak_adsorptiondesorption_2013}. Trace water in the gas phase further promotes hydroxide formation (Cu(OH)$_2$), which participates in carbonate and bicarbonate production. These results establish the dual physical and chemical character of CO$_2$ adsorption on oxidized copper surfaces and highlight the essential role of hydroxyl groups as precursors to surface carbonate growth.
        
        Theoretical and \textit{in situ} studies later refined this picture by isolating the role of subsurface oxygen and electric fields. Garza \textit{et al.}~\cite{garza_is_2018} demonstrated through DFT calculations that subsurface oxygen beneath the second Cu(111) layer only weakly perturbs CO$_2$ physisorption and provides no significant stabilization for chemisorbed species. The authors concluded that subsurface oxygen is not required for CO$_2$ reduction, and that the enhanced reactivity of oxide-derived copper likely stems from surface defects and under-coordinated sites generated during oxide reduction. Consistent with this, Fields \textit{et al.}~\cite{fields_role_2018} showed that oxygen trapped within the first three atomic layers of Cu diffuses rapidly under reducing conditions, implying that the observed catalytic effects are due to near-surface rather than bulk oxygen.
        
        An additional key factor is the influence of external electric fields. Jafarzadeh \textit{et al.}~\cite{jafarzadeh_activation_2020} found that electric fields above 0.1~V\,\AA$^{-1}$, in combination with surface roughness and excess electrons, can convert the weakly physisorbed CO$_2$ configuration into a chemisorbed, bent carbonate-like intermediate, facilitating C–O bond elongation and activation. Rough or defect-rich Cu facets such as Cu(211) and Cu(110) respond more effectively to these fields, leading to stronger binding and higher activation degrees than flat Cu(111). These results unify electrochemical and plasma-catalytic perspectives by linking CO$_2$ activation directly to the interplay of morphology, field strength, and surface charge.
        
        Together, these studies delineate a coherent framework for CO$_2$–Cu interactions: (i) metallic copper primarily supports weak physisorption of linear CO$_2$; (ii) partial oxidation and surface hydroxylation induce the formation of carbonate and bicarbonate species; (iii) subsurface oxygen plays a negligible direct role under reducing potentials; and (iv) strong electric fields and morphological irregularities can locally trigger CO$_2$ activation into bent, chemisorbed configurations. These principles underpin the understanding of both catalytic and non-catalytic phenomena involving CO$_2$–Cu interfaces, including surface charging and degradation processes in copper-based devices and detectors.

        These principles establish the conceptual baseline for interpreting the NAP-XPS spectra obtained on GEM copper electrodes under CO$_2$ exposure.

\section{Results}

    A series of NAP-XPS measurements was conducted to investigate the interaction between the copper electrode surface of GEM foils identical to those used in the ALICE experiment, but not operated in it, and carbon dioxide gas. The samples were prepared as described in section~\ref{sec:sample_prep}. 
    
    \subsection{Analysis of Cu 2p signals}
    
\begin{figure*}[htb!]
            \centering
            \includegraphics[width=1.0\linewidth]{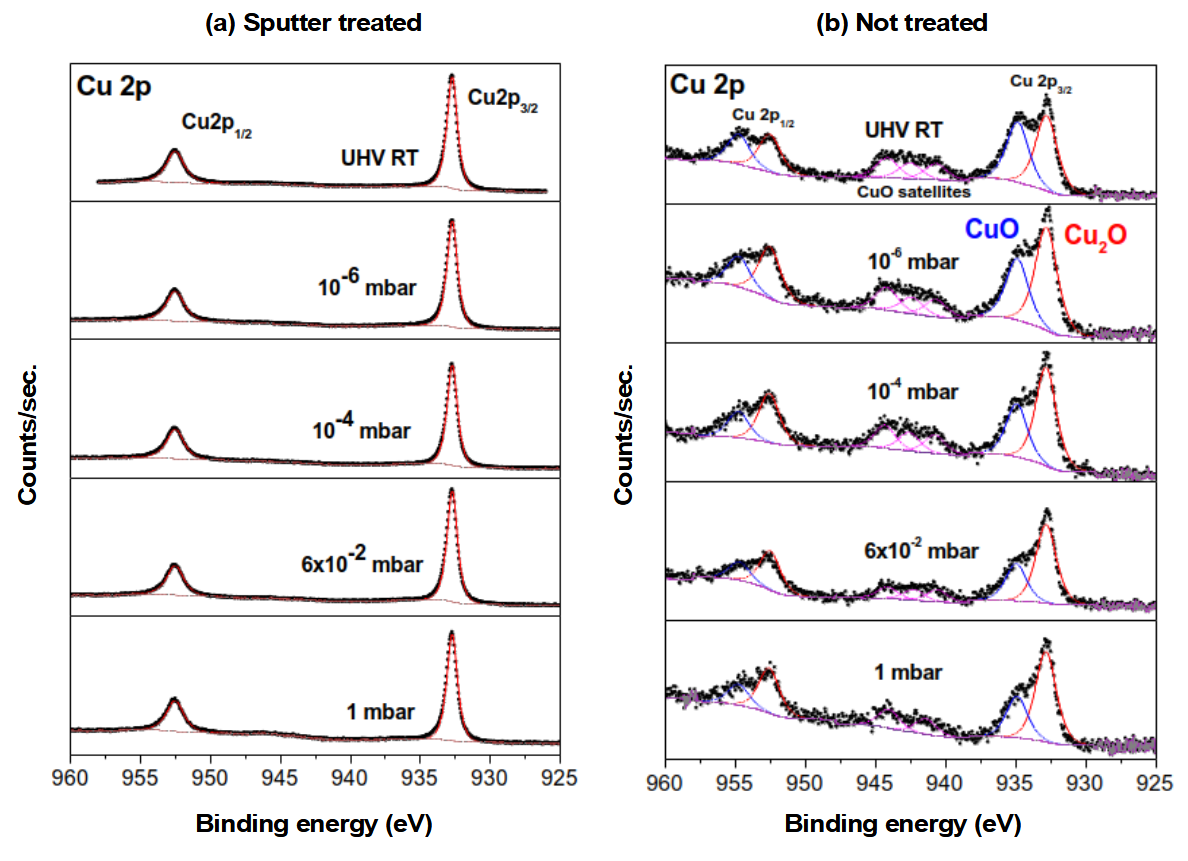}
            \caption{Binding energy spectra obtained by NAP-XPS for the Cu~2p core levels under different CO$_2$ pressures. On the left, the metallic copper (Cu$^0$) signal remains unchanged for the sputtered sample. On the right, CuO (Cu$^{2+}$) and Cu$_2$O (Cu$^+$) components are shown for the untreated sample, with progressive attenuation of the CuO peak at higher pressures.}
            \label{fig:napxps-cobre}
        \end{figure*}

        The Cu~2p core-level spectra were analyzed to evaluate the oxidation states of copper under different CO$_2$ pressures for both surface conditions. For the sputter-cleaned surface, the Cu~2p$_{3/2}$ spectrum was dominated by a metallic component (Cu$^0$) at 932.6~eV, characterized by the absence of shake-up satellite features, indicating the stability of the metallic phase under $CO_2$ pressure, see Figure~\ref{fig:napxps-cobre}-(a).
        
\begin{figure*}[htb!]
            \centering
            \includegraphics[width=0.75\linewidth]{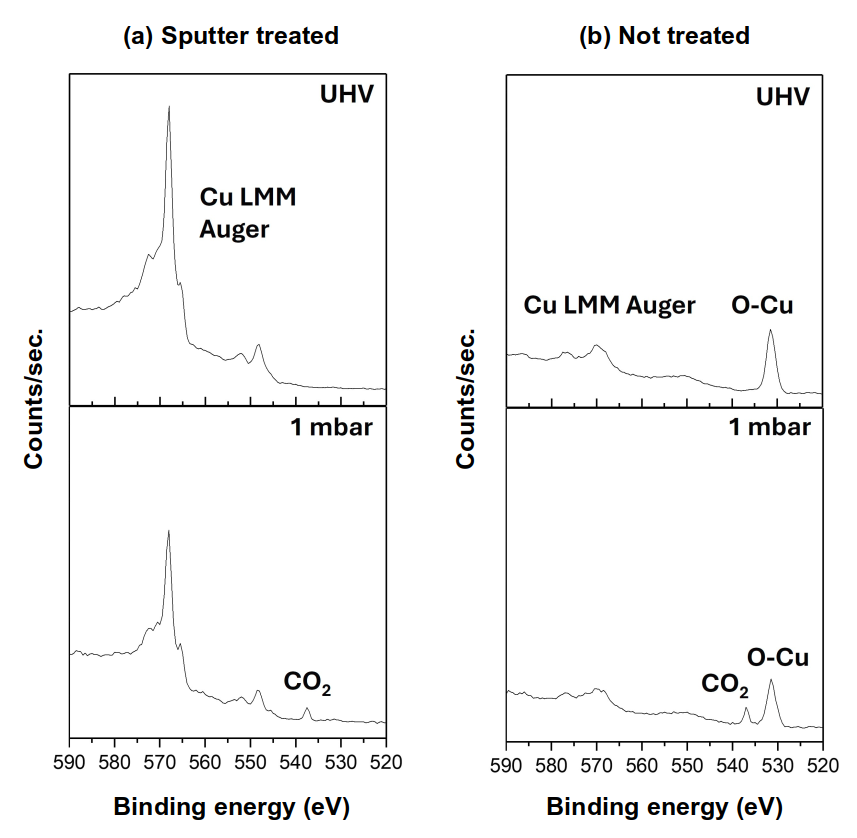}
            \caption{The Cu LMM Auger transitions and the O 1s XPS regions obtained by NAP-XPS for the lowest and highest tested CO$_2$ pressures comparing the sputter-cleaned (left) and untreated (right) samples. The Auger features of the cleaned surface retain a metallic signature, while the untreated surface exhibits oxide characteristics.}
            
            \label{fig:o1s}
        \end{figure*}

        In contrast, the untreated samples exhibited spectral components corresponding to Cu$_2$O (Cu$^+$, 932.7~eV) and CuO (Cu$^{2+}$, 934.9~eV). The CuO component showed a gradual decrease in intensity as the CO$_2$ pressure increased, suggesting partial reduction of the oxidized copper surface in the presence of CO$_2$, see Figure~\ref{fig:napxps-cobre}(b). The overlapping binding energies of Cu$^0$ and Cu$^+$ were distinguished using the Cu LMM Auger transitions (in the O 1s region), which confirmed that metallic copper was present only in the sputtered samples, while the untreated surfaces remained oxidized. See Figure~\ref{fig:o1s}.  
        
        Fig.~\ref{fig:o1s} also exhibits a weak but distinct peak associated with CO$_2$, corresponding to ionized molecular species generated in the near-surface region. The observation of ionized CO$_2$ species in the O 1s region suggests that part of the gas phase is ionized in the vicinity of the surface during acquisition. While the ionization mechanism differs from the electron avalanches occurring in GEM detectors, this observation is relevant because CO$_2^{+}$ and other ionized fragments produced during detector operation can also reach and interact with the copper surfaces. 
        
\begin{figure}[htb!]
            \centering
            \includegraphics[width=1.0\linewidth]{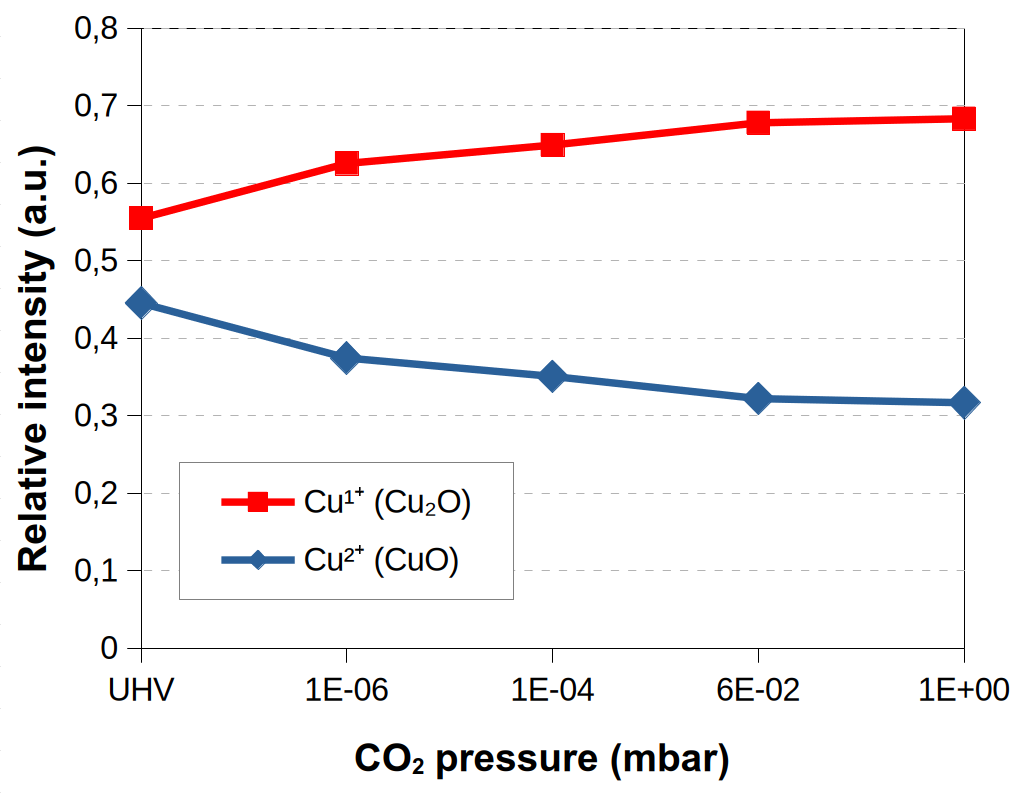}
            \caption{Relative intensities of the Cu~2p$_{3/2}$ components for untreated samples. The trend indicates a decrease in the CuO fraction relative to Cu$_2$O as the CO$_2$ pressure increases.}
            \label{fig:cu_tendencias}
        \end{figure}

        The relative intensities of the Cu~2p$_{3/2}$ components were extracted from fitted spectra using the CasaXPS software. A consistent decrease of the CuO signal relative to Cu$_2$O was observed with increasing CO$_2$ pressure until reaching a stable ratio near 3:7, as summarized in Figure~\ref{fig:cu_tendencias}. This trend supports the interpretation that CO$_2$ interacts preferentially with the more oxidized Cu$^{2+}$ species, leading to a mild surface reduction.
        
    \subsection{Analysis of C 1s signals}
    
\begin{figure*}[htb!]
            \centering
            \includegraphics[width=1.0\linewidth]{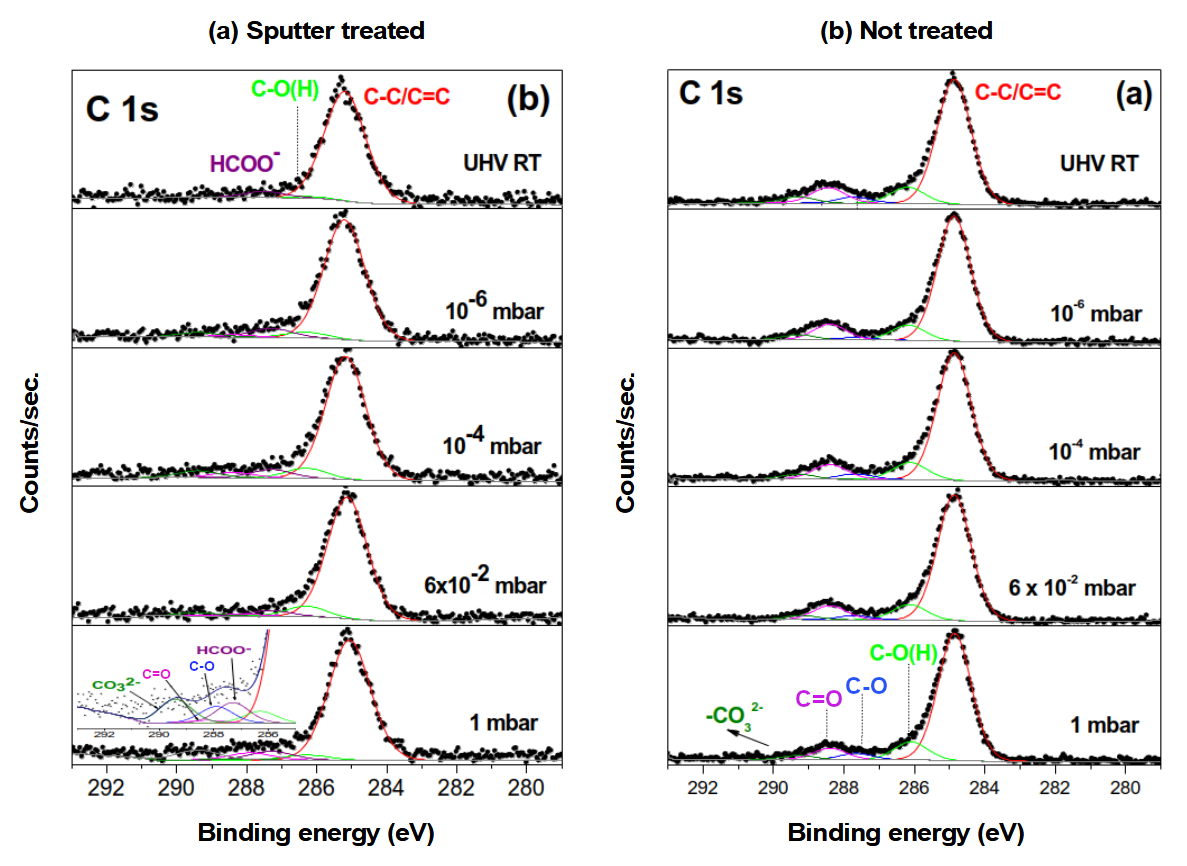}
            \caption{C~1s photoelectron spectra for the sputter-cleaned (a) and untreated (as delivered) (b) samples under different CO$_2$ pressures. The untreated sample exhibits distinct contributions from C=O (288.4~eV), C-O (287.9~eV), CO$_3^{2-}$ (289.4~eV), and C–O(H) (286.3~eV).}
            \label{fig:napxps-carbono}
        \end{figure*}

        The C~1s region was analyzed to identify the nature of carbon species adsorbed or formed on the GEM surface during CO$_2$ exposure. For the sputter-cleaned copper surface, the spectra were dominated by the adventitious carbon component (C–C/C–H at 284.8~eV) with no detectable evolution under varying CO$_2$ pressures, confirming the absence of significant adsorption or reaction (Figure~\ref{fig:napxps-carbono}-(a)). 
        
        The untreated samples, however, displayed multiple contributions whose relative intensities evolved with gas pressure. These included carbonyl (C=O, 288.4~eV), C-O (287.9~eV) bonds, carbonate species (CO$_3^{2-}$, 289.4~eV), and hydroxylated carbon (C–O(H), 286.3~eV), as shown in Figure~\ref{fig:napxps-carbono}-(b). The fitting parameters and peak assignments followed standard energy values from the literature~\cite{favaro_subsurface_2017}.  
        
\begin{figure}[htb!]
            \centering
            \includegraphics[width=1.0\linewidth]{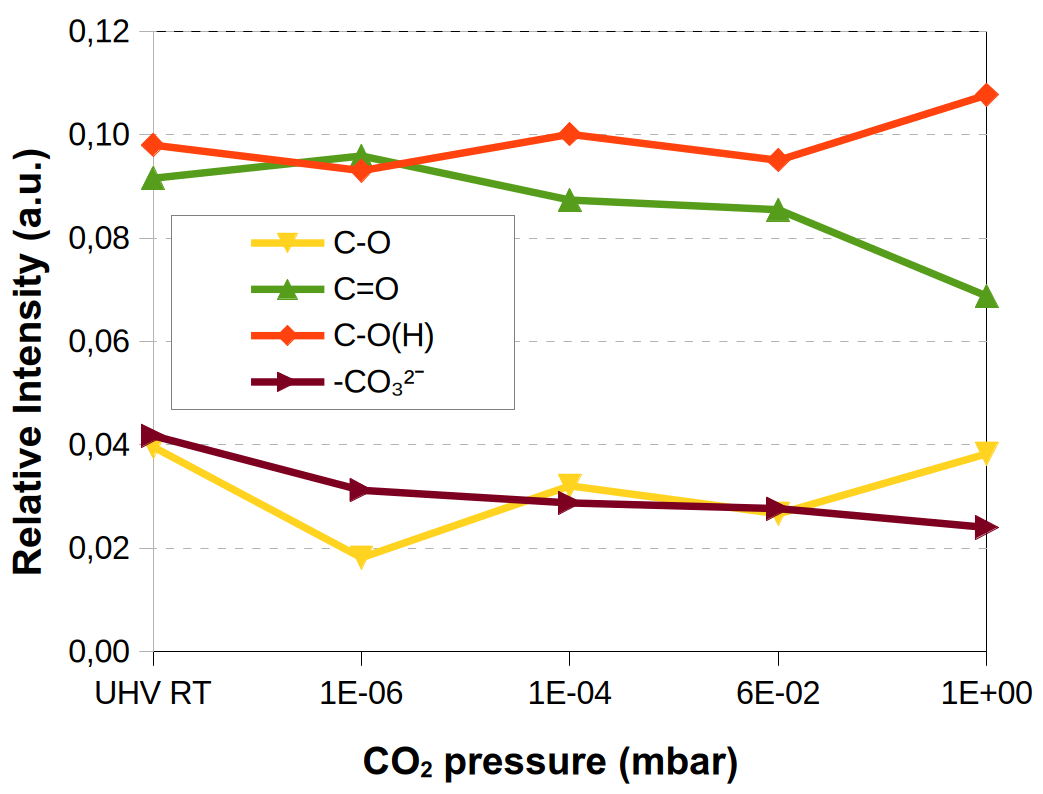}
            \caption{Relative intensities of the C~1s components for the untreated sample. The plotted ratios correspond to the fitted peak areas normalized to the total C~1s intensity.}
            \label{fig:c_tendencias}
        \end{figure}

        The evolution of these components with pressure, presented in Figure~\ref{fig:c_tendencias}, indicates competition between physisorption, chemisorption, and surface hydroxylation processes on oxidized copper surfaces. The consistency of peak positions and widths across measurements ensured that all variations in intensity reflected genuine chemical changes rather than instrumental artifacts.
         
    \subsection{Surface distribution of oxidation states}

\begin{figure}
            \centering
            \includegraphics[width=1\linewidth]{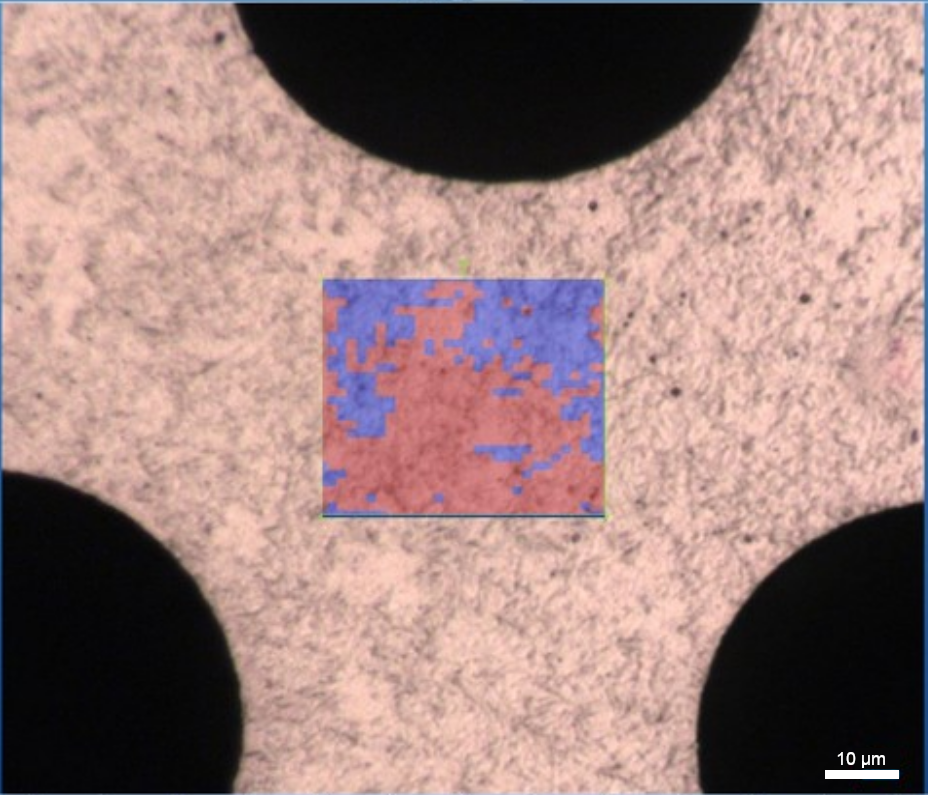}
            \caption{Optical micrograph of the GEM foil with Raman mapping overlay. The Raman spectroscopy mapping of the GEM copper surface (30 × 30 points) shows the spatial distribution of copper oxide phases. The intensity map of the band centered at ~150 cm$^{-1}$ (red) corresponds to Cu$_2$O, while the band near ~300 cm$^{-1}$ (blue) is attributed to CuO. The results indicate a predominantly Cu$_2$O surface with spatially heterogeneous contributions from CuO.}
            \label{fig:raman}
        \end{figure}

        Raman spectroscopy was employed to identify and spatially resolve copper oxide phases present on the GEM foil surface. Figure \ref{fig:raman} shows a Raman-based spatial map obtained from a 30 × 30 point discretization, in which each measurement point is assigned to the dominant oxide phase based on the relative intensity of the characteristic vibrational bands of copper oxides.

\begin{figure}[htb!]
            \centering
            \includegraphics[width=1\linewidth]{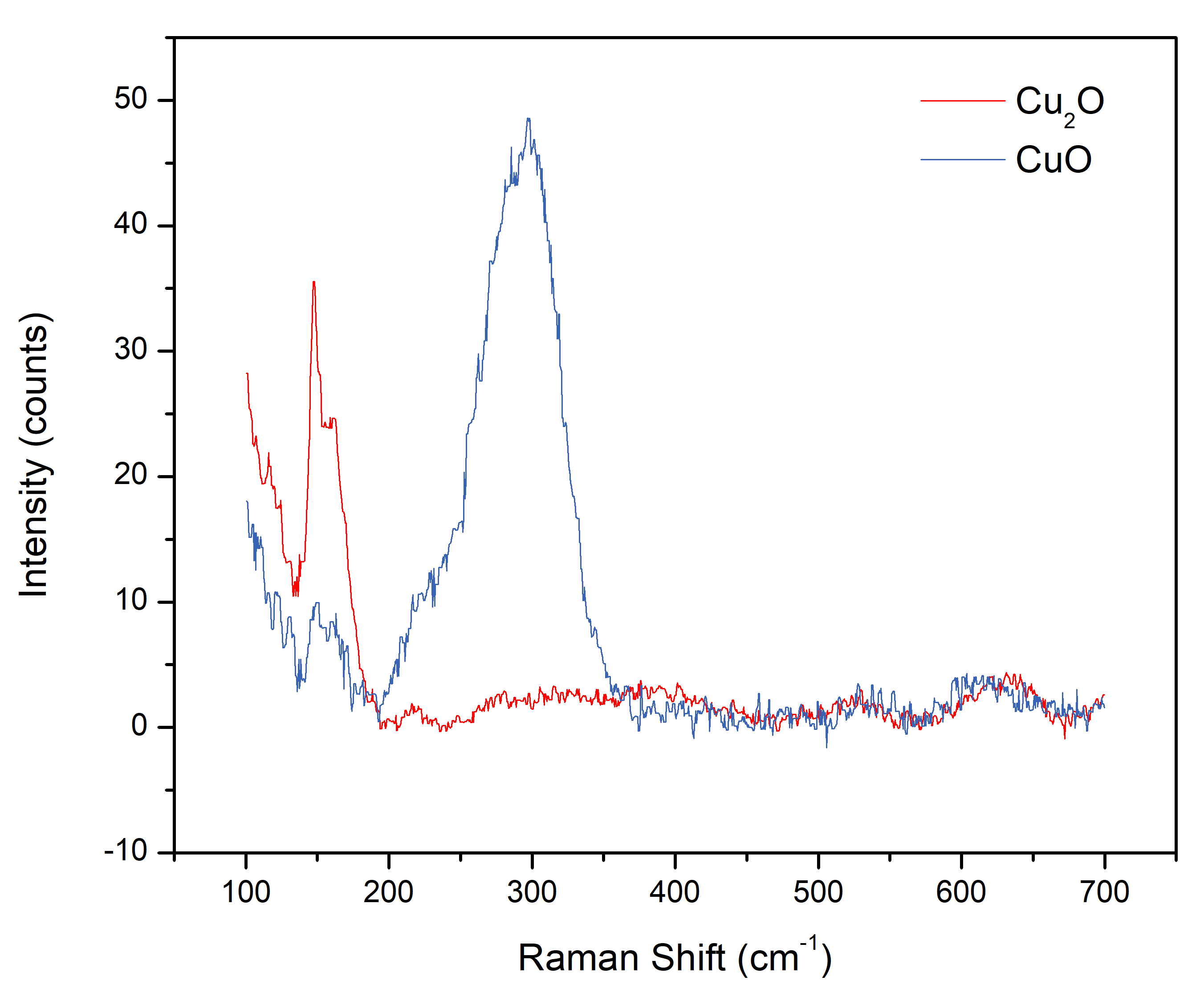}
            \caption{Typical spectra observed during surface analysis with Raman spectroscopy. }
            \label{fig:raman_spectra}
        \end{figure}

        Two dominant spectral features were identified across the analyzed region, as presented in two typical spectra in Fig.~\ref{fig:raman_spectra}. A band centered at approximately 150 cm$^{-1}$, attributed to Cu$_2$O, was broadly distributed over the mapped area. In addition, a band near 300 cm$^{-1}$, characteristic of CuO, was also detected, though with comparatively lower and spatially heterogeneous intensity. The coexistence of these two Raman-active modes indicates that the surface of the as-delivered samples (untreated) is composed predominantly of Cu$_2$O with localized contributions from CuO.

        The Raman mapping reveals a non-uniform spatial distribution of copper oxide phases, suggesting that the surface oxidation state is heterogeneous at the micrometer scale. This heterogeneity may play a role in governing local surface reactivity and charge accumulation processes relevant to GEM operation.

\section{Discussion}

    The NAP-XPS measurements presented here link the evolution of copper oxidation states and carbon‐containing surface species to well‐established principles of CO$_2$–Cu chemistry, while providing detector‐relevant context for GEM aging. First, our Cu~2p analysis shows that the sputter‐cleaned surface remains metallic (Cu$^0$) within the explored pressures, whereas untreated foils display Cu$^+$ and Cu$^{2+}$ components that evolve under CO$_2$. This behavior is consistent with the literature picture in which metallic Cu weakly interacts with linear CO$_2$, while partially oxidized surfaces and hydroxylated sites stabilize carbonate/bicarbonate‐type species~\cite{regoutz_influence_2018, isahak_adsorptiondesorption_2013, favaro_subsurface_2017}. The minute energy separation between Cu$^0$ and Cu$^+$ at 2p$_{3/2}$ makes O 1s corroboration essential, and resolves the ambiguity, in line with standard XPS practice~\cite{Moulder1992}.
    
    In the C~1s region, the assignment of contributions at $\sim$286.3~eV (C–O(H)), $\sim$287.9~eV (C-O), $\sim$288.4~eV (C=O), and $\sim$289.4~eV (CO$_3^{2-}$) follows the binding‐energy ranges reported for CO$_2$ adsorption on copper and copper oxides~\cite{favaro_subsurface_2017, regoutz_influence_2018, isahak_adsorptiondesorption_2013}. These components reflect the central role of surface states: partial oxidation and local charge/field effects favor bent intermediates and carbonate formation, whereas clean Cu sustains mainly weak physisorption. Theory further supports that subsurface oxygen is not a prerequisite for CO$_2$ activation on Cu, and that field/morphology can amplify the binding of activated (C-O) configurations~\cite{garza_is_2018, fields_role_2018, jafarzadeh_activation_2020}. Altogether, our spectra align with a scenario where CO$_2$ reacts primarily at oxidized surfaces, producing oxygenated layers whose electrical properties will depend on the balance among hydroxyl, carbonate, and oxide species.

    The Raman results confirm the coexistence of Cu$_2$O and CuO phases, with a dominant contribution from Cu$_2$O, in agreement with the relative intensities extracted from the Cu 2p analysis. Importantly, the Raman maps reveal a spatially heterogeneous distribution of these oxides at the micrometer scale. Such heterogeneity indicates that the surface oxidation state is not uniform but rather composed of domains with distinct local chemical environments. This observation supports the interpretation that CO$_2$ activation and carbonate formation are likely governed by localized oxidized sites rather than by a homogeneous surface layer.
    
    A second point of discussion concerns operational relevance. The observation of a CO$_2$ feature in the O 1s window indicates the ionization of gas‐phase molecules in the near‐surface region during acquisition, providing a controlled environment in which similar molecular ionization pathways may occur. This is advantageous for bridging surface spectroscopy to detector behavior. It also suggests that near-surface ionization processes may influence surface charge accumulation mechanisms. Besides, it connects with recent evidence that aging in GEMs can arise from a competition between deposition and etching chemistry in high‐field microenvironments, including polymer redeposition and interfacial element migration observed in our group’s work~\cite{saramela_evidence_2024}, and geometry/roughness changes reported at very high doses or in CF$_4$‐rich conditions~\cite{poli_lener_irradiation_2024}.

    While the present study does not directly measure aging rates in operating detectors, it provides surface-chemical evidence consistent with the improved stability observed in CO2-based GEM operation. In this context, our results help rationalize why CO2-based mixtures tend to be less aggressive than hydrocarbon or strongly reactive chemistries for long-term GEM operation. Classical and modern studies point out that aging is exacerbated by polymerizable impurities (notably Si‐bearing species) and by conditions that favor persistent radical growth; conversely, CO$_2$ as an inorganic quencher and appropriate gas refresh mitigate polymerization‐driven deposits~\cite{sauli_fundamental_2003, vavra_physics_2003, niebuhr_aging_2006, procureur_why_2020, corbetta_triple-gem_2020}. Within this framework, the thin inorganic films detected under CO$_2$ (hydroxyl/carbonate/oxide) may behave more like thin, partially passivating layers than like aggressively insulating carbonaceous polymers, provided that contaminants are controlled and CF$_4$ or hydrocarbon fragments are minimized~\cite{procureur_why_2020, corbetta_triple-gem_2020}. Hence, the combined picture that emerges is: (i) surface state and trace impurities govern the chemical identity of the emerging film; (ii) CO$_2$ alone biases the system toward comparatively milder oxygenated species; and (iii) detector lifetime is therefore strongly coupled to gas and surface purity, materials selection, and operating fields, as already emphasized in aging surveys and targeted GEM studies~\cite{sauli_gem:_1997, sauli_fundamental_2003, fallavollita_advanced_2020, niebuhr_aging_2006}.

    To our knowledge, this work provides one of the first direct spectroscopic investigations linking the surface chemistry of copper electrodes to their interaction with CO$_2$ under conditions relevant to GEM detectors. These observations offer an experimental surface-chemical perspective that helps rationalize the empirically observed robustness of CO$_2$-based gas mixtures in GEM operation. 

\section{Conclusions}

    The NAP-XPS and complementary Raman analyzes presented here provide a coherent picture of the initial chemical and structural interactions between CO$_2$ and the copper surfaces that compose GEM electrodes. The experiments demonstrate that CO$_2$ exposure promotes a partial reduction of CuO to Cu$_2$O in untreated samples, while sputter-cleaned foils remain metallic and chemically stable. These results indicate that the surface oxidation state is a key factor in defining how CO$_2$ interacts with copper, supporting the notion that the presence of oxygen and hydroxyl species facilitates the formation of carbonate and bicarbonate groups. 

    The Raman mapping further confirms that the surface is predominantly composed of Cu$_2$O with spatially heterogeneous contributions from CuO, providing structural support for the oxidation-state evolution inferred from XPS. The micrometer-scale heterogeneity revealed by Raman suggests that local variations in oxide distribution may influence site-specific reactivity and charge stabilization mechanisms under detector operation.
    
    The C~1s spectra revealed that these oxygenated species coexist with C=O and CO. Together, the copper and carbon spectral evolutions suggest that CO$_2$ can interact with oxidized copper surfaces and actively participate in mild reduction and surface passivation pathways. The oxygenated films formed under these conditions appear to be thin and predominantly inorganic, contrasting with the thick polymeric or carbonaceous deposits typically observed in detectors operated with hydrocarbon or CF$_4$-based mixtures.
    
    These findings reinforce the understanding that the long-term stability of GEM detectors operated in CO$_2$ mixtures arises from the gas’s ability to establish reversible equilibria with the electrode surface, favoring the formation of self-limiting oxygenated layers rather than irreversible insulating films. Such oxygenated layers are expected to be less prone to charge accumulation than the polymeric or carbonaceous deposits typically associated with hydrocarbon-based mixtures, which may contribute to the improved long-term stability observed in CO$_2$-based GEM operation. The controlled ionization of CO$_2$ molecules near the surface, observed in O 1s spectra, further supports that the measurements capture relevant aspects of the gas–surface interactions occurring in operating GEM detectors. 

    The Raman mapping further reveals that the copper oxide phases are not uniformly distributed across the GEM surface but rather form micrometer-scale domains where Cu$_2$O and CuO coexist with different relative intensities. This spatial heterogeneity indicates that the chemical state of the copper surface is intrinsically non-uniform even before detector operation. Such variations in local oxidation states may lead to spatial differences in surface reactivity, adsorption behavior, and charge accumulation properties. In the context of GEM detectors, this chemical heterogeneity may contribute to the non-uniform patterns of material redeposition and aging phenomena often observed under irradiation~\cite{saramela_evidence_2024}, where localized surface chemistry can influence the initiation and growth of deposits. These observations also suggest that controlling the initial oxidation state of GEM copper electrodes may be an effective strategy to mitigate aging phenomena in CO$_2$-based detector operation.
    
    By connecting these surface-level observations with established models of detector degradation~\cite{vavra_physics_2003, sauli_fundamental_2003, niebuhr_aging_2006, fallavollita_advanced_2020, poli_lener_irradiation_2024}, this work provides experimental evidence that complements and strengthens ongoing aging studies of the ALICE-TPC GEM system. The results confirm that both chemical and morphological stability under CO$_2$-based operation is closely linked to the intrinsic reactivity of the copper surface and to the benign nature of the oxygenated products formed in this environment. These findings are also relevant to GEM detectors more broadly, as CO$_2$-based gas mixtures are widely employed across different applications.

\section*{Acknowledgments}

The authors thank the financial support provided by INCT-CERN-Brasil (project number 406672/2022-9), by CNPq (project numbers 406982/2021-0,  305793/2021-7 306414/2022-8, and 300313/2025-0), and by FAPESP (project numbers 2020/04867-2, 2022/03043-1, and  2018/19240-5).

\hfill \break
During the preparation of this work, the authors used ChatGPT by OpenAI to improve language and readability. After using this tool, the authors reviewed and edited the content as needed and assume full responsibility for the content of the publication.





\bibliographystyle{model1-num-names}
\bibliography{references.bib}







\end{document}